\documentclass[10pt,letterpaper,typeset,superscriptaddress,twocolumn,showpacs]{revtex4}

\usepackage{amsfonts}
\usepackage{amsmath}
\usepackage{amssymb}
\usepackage{graphicx}
\usepackage{textcomp}
\usepackage{float}
\usepackage[normalem]{ulem}

\begin{document}

\title{Minimum tomography of two entangled qutrits using local measurements of one-qutrit SIC-POVM}

\author{W. M. Pimenta}
\email{wmpfis@fisica.ufmg.br}
\author{B. Marques}
\author{T. O. Maciel}
\author{R. O. Vianna}
\affiliation{Departamento de F\'{\i}sica, Universidade Federal de Minas Gerais, Caixa Postal 702, Belo~Horizonte,~MG 30123-970, Brazil.}
\author{A. Delgado}
\author{C. Saavedra}
\affiliation{Center for Optics and Photonics, Universidad de Concepci\'on, Casilla 4016, Concepci\'on, Chile.}
\affiliation{Departamento de F\'{\i}sica, Universidad de Concepci\'on,  Casilla 160-C, Concepci\'on, Chile.}
\author{S. P\'adua}
\affiliation{Departamento de F\'{\i}sica, Universidade Federal de Minas Gerais, Caixa Postal 702, Belo~Horizonte,~MG 30123-970, Brazil.}

\begin{abstract}
An experimental demonstration of two-qutrit state tomography via one-qutrit symmetric informationally complete positive operator-valued measure (SIC-POVM) is presented. A two-qutrit state is encoded in the transversal paths of a spontaneous parametric down-converted photon pair. A spatial light modulator allows to implement the necessary 81 POVM elements to reconstruct the state. The quality of the reconstruction is evaluated by comparing independent measurements with the predicted results calculated with the reconstructed state. Entanglement in the system is calculated via negativity and generalized robustness from the two-qutrit reconstructed density operator.
\end{abstract}

\date{\today}
\maketitle


Qudits or $d$-dimensional quantum systems have been the focus of an intensive research effort in the last decade because of possible applications and challenges in quantum information
theory \cite{caves,genovese}. Higher dimensional quantum states are useful in quantum communications because more information can be encoded in only one system by using higher alphabets when compared with two-dimensional ones \cite{peres}. Quantum key distribution with $d$-dimensional quantum states can be safer \cite{peres, durt, bruss, bourennane, cerf}. Maximally entangled qudits have been shown to produce violations of local realism stronger than maximally entangled qubits \cite{kaszlikowski}. Generalized Bell inequalities for qudits were found \cite{collins, thew2} and their violation has been used as a test of entanglement for qudits systems \cite{dada}. Quantum computation by using qudits has been proposed \cite{qucoqudi} and quantum information tasks as cloning \cite{fabio}, entanglement concentration \cite{vaziri}, quantum key distribution \cite{groblacher, thew1, bogdanov, walborn1}, quantum game demonstration \cite{game} and quantum bit commitment protocol \cite{langford} have been demonstrated.
Violation of local realism for qudits is shown to grow with the dimension d \cite{massar}, and smaller quantum efficiency for the detectors are required for detection loophole free Bell tests \cite{vertesi}. Qutrit states have been  experimentally generated in several degrees of freedom of two photons such as polarization \cite{polar1, polar2, polar3, polar4}, orbital angular momenta \cite{vaziri, mair, langford, dada, ang5}, time-bin \cite{thew1, thew2}, and transversal path \cite{trans1, trans2, trans3}

In general manipulating (or controlling) higher dimensional quantum states is not a simple task because for many systems it is difficult to implement the necessary unitary transformations or the measurement operators. Two important and particularly difficult problems are the assessment of the quality of experimentally generated higher dimensional quantum states and of the determination of the entanglement degree. A few works have measured entanglement witnesses for two qudits systems entangled in orbital angular momentum \cite{boyd4} or transversal path \cite{chile, marcos, alejandra}. An effective way to obtain information about a quantum system is to measure its density operator, i.e., to perform the quantum state tomography \cite{fano, james}. This technique has been used for measuring the density operator of photonic qudits in orbital angular momenta \cite{langford, agnew}, time bin \cite{timbito}, and transversal path \cite{trans3, path2}. A drawback of this technique is the high number of measurements
necessary for obtaining the density operator of the prepared state and the consequent increase of experimental error propagation when solving the set of equations necessary to obtain the elements of the density operator. In
reference \cite{agnew}, they have made 14400 measurements for reconstruct the density operator of a two-qudit system (d=8), well above the minimum required number of measurements: 4096. It becomes clear then that minimum quantum tomography schemes that produce an accurate reconstruction of the density operator are very important to characterize higher dimensional systems \cite{rehacek, cingapura, wander, englert}.

Two optimal methods to perform quantum state tomography are the ones based on mutually unbiased bases (MUBs) and symmetric informationally complete positive operator-valued measure (SIC-POVM). In the MUBs case, bases have the property to be maximally incompatible. In other words, a state producing precise measurement results in one base produces maximally random results in all the others \cite{Klimov}. Quantum tomography using this scheme was implemented in \cite{path2} for the state reconstruction of photons entangled in the transversal momentum  degree of freedom. On the other hand, a SIC-POVM is composed of rank-one operators such that the product between any two of them is a constant number \cite{caves2, flammia, flammia2}. Mendendorp \textit{et al.} performed a one-qutrit SIC-POVM, encoding the qutrit in polarization states of a photon with two spatial modes \cite{medendorp}. This last scheme is important for being part of a largest group of
minimum measurements tomography schemes, specially useful in high dimensional systems as discussed above. No SIC-POVM experimental tomographic characterization of high dimensional bipartite system, i. e., two qudits entangled states, have been demonstrated. A recent experimental scheme for implementing SIC-POVM quantum tomography using multiport devices in path encoded qutrits have been proposed by N. M. Tabia \cite{tabia}.

Spontaneous parametric down conversion (SPDC) is a natural source of correlated photon pairs and has been the most used photons source in quantum optics studies \cite{Ou}. Photons generated by SPDC are easy to
manipulate \cite{Chuang} and can be correlated in different degrees of freedom such as polarization
\cite{Ou, Shih}, transversal \cite{trans1}, longitudinal \cite{Rarity, Rossi} and orbital momenta \cite{mair}, energy-time \cite{Franson}, and time
\cite{Brendel,Rossi2}, or even in more than one degree of freedom when are prepared in hyper entangled states \cite{Cinelli, Yang, Santos}.
To implement the SIC-POVM in the transversal path degree of freedom is necessary to have means of controlling the phases and amplitudes of the elements of the POVM. In \cite{Lima}, it is shown that this can be done by using a Spatial Light Modulator (SLM). The SLM has been used for tomographic reconstruction of photonic states entangled in polarization \cite{Cialdi}, transversal \cite{wander} and orbital angular momenta \cite{agnew}. Moreover, the SLM has been used to demonstrate a Bell's inequality violation in photons prepared in orbital angular momentum state \cite{Leach} and minimum Deutsch quantum algorithm \cite{breno}.

A SLM is used in this work to implement the SIC-POVM elements necessary to reconstruct the density operator of two photonic spatial qutrits. Two qutrits entangled in transversal path are produced when two photons generated by SPDC cross a triple-slit. Sanchez \textit{et al.} obtained analytically in \cite{Sanchez}, a specific SIC-POVM for odd dimensions. Here, the experimental tomography of the photonic two-qutrit entangled state in transversal path using a one-qutrit SIC-POVM is demonstrated. The density operator of the prepared entangled photons in spatial variables is obtained and the degree of entanglement of the two-qutrit system is calculated.

This paper is organized as follows: In section 1, we give explicit expressions for states involved in SIC-POVM for qutrits. The experimental results are presented in section 2. In section 3, we discuss the quality of the reconstructed density operator. The distance between the predicted prepared state and the reconstructed one is calculated. A measured two-photon interference pattern is compared with the predicted interference pattern calculated from the reconstructed density operator. Entanglement of the prepared two-qutrit system is also calculated from the obtained density operator for a different number of POVM elements. This work ends with conclusions in section 4.

\section{Theory}

In \cite{Sanchez}, it was calculated the SIC-POVM suitable for realizing the quantum tomography of a qutrit state. The states that generate the SIC-POVM elements are equidistant. They are nonorthogonal states  $|\psi_{j}\rangle$ for the inner product between any two equidistant states is equal to a complex number or its conjugate, which means
$\langle \psi_{j}| \psi_{j'}\rangle=\alpha=|\alpha|e^{i\theta}$, $\forall$ $j>j'$. Moreover, when the inner product is real and positive (equal to $1/\sqrt{d+1}$) and the sum of the POVM elements
is equal to the identity we have a special type of POVM: the so-called SIC-POVM. For one-qutrit state the nine measurement operators in the
$\{|0\rangle, |1\rangle, |2\rangle\}$ basis are \cite{Sanchez}:

\begin{equation}
 \hat{\Pi}_{j}=\frac{1}{18}|\psi_{j}\rangle \langle \psi_{j}| \label{1}
\end{equation}
where $|\psi_{j}\rangle$ with $j=1,..,9$ is defined as:
\begin{eqnarray}
|\psi_{1} \rangle = \frac{1}{\sqrt{2}} (| 0 \rangle + | 2 \rangle),\hspace{20mm} \nonumber
\\
|\psi_{2} \rangle = \frac{1}{\sqrt{2}} (e^{2\pi i/3}| 0 \rangle + e^{-2\pi i/3} | 2 \rangle), \nonumber
\\
|\psi_{3} \rangle = \frac{1}{\sqrt{2}} (e^{-2\pi i/3}| 0 \rangle + e^{2\pi i/3} | 2 \rangle), \nonumber
\\
|\psi_{4} \rangle = \frac{1}{\sqrt{2}} (| 1 \rangle + | 0 \rangle), \hspace{20mm} \nonumber
\\
|\psi_{5} \rangle = \frac{1}{\sqrt{2}} (e^{2\pi i/3}| 1 \rangle + e^{-2\pi i/3} | 0 \rangle),  \label{2}
\\
|\psi_{6} \rangle = \frac{1}{\sqrt{2}} (e^{-2\pi i/3}| 1 \rangle + e^{2\pi i/3} | 0 \rangle), \nonumber
\\
|\psi_{7} \rangle = \frac{1}{\sqrt{2}} (| 2 \rangle + | 1 \rangle), \hspace{20mm} \nonumber
\\
|\psi_{8} \rangle = \frac{1}{\sqrt{2}} (e^{2\pi i/3}| 2 \rangle + e^{-2\pi i/3} | 1 \rangle), \nonumber
\\
|\psi_{9} \rangle = \frac{1}{\sqrt{2}} (e^{-2\pi i/3}| 2 \rangle + e^{2\pi i/3} | 1 \rangle),  \nonumber
\end{eqnarray}
and as mentioned above $\sum_{1}^{9}\hat{\Pi}_{j}=1$.

This optimal SIC-POVM calculated for the one-qutrit quantum tomography can be used for implementing a minimum quantum state tomography of a bipartite two-qutrit system. For this case the POVM needed for reconstructing the two-qutrit state is the set $\{\Pi_{i, j}\}$, $\Pi_{i, j}=\Pi_{i}\otimes \Pi_{j}$ generated by the above states and forming a set of 81 elements. Notice that the POVM set $\Pi_{i,j}$ is not SIC for the two-qutrit bipartite state system although is informationally complete.

\section{Experimental Setup}

\begin{figure}[!htbp]
		\includegraphics[width=8.5cm]{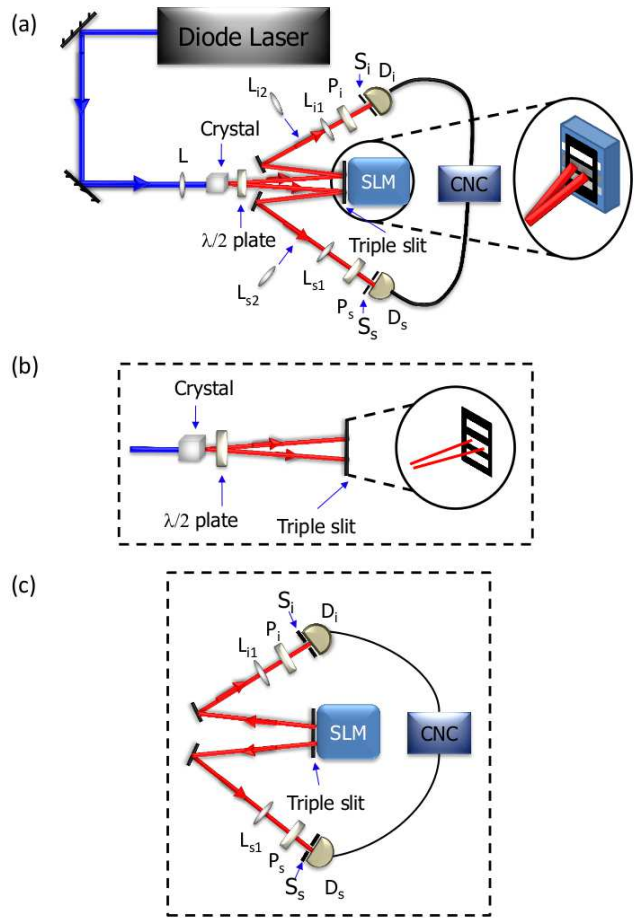}
		\vspace{0.0cm}
		\caption{\small{(Color online) (a) Experimental setup scheme for the two-qutrit minimum quantum state tomography. L lens focuses the pump beam at the triple-slit plane;
		lenses L$_{s1}$ and L$_{i1}$ are used to detect the signal
                and idler reflected beams at the Fourier plane, while lenses L$_{s2}$ and L$_{i2}$ are used to project the triple slits image at the detectors'
		planes. A half-wave plate is placed right after the crystal and polarizers P$_{i}$ and P$_{s}$ are positioned in front of APDs detectors.
		CNC is a coincidence counter and SLM is the Spatial Light Modulator. (b) State preparation part of the setup. (c) State determination part of the setup.}}\label{montagem}
\end{figure}

The experimental setup scheme is shown in Fig. \ref{montagem}. A 50 mW Diode Laser operating at 405 nm is used to pump a 2-mm-thick $BiB_3O_6$ ($BIBO$) crystal and generates, by type I SPDC, degenerate non-collinear photon pairs. Signal and idler ($\lambda_{s,i}$=810 nm) beams pass through a $\lambda/2$ plate
(half-wave plate) before they cross a triple-slit placed perpendicular to the signal and idler beams direction. The triple-slit is placed at a distance of 250 mm from the crystal. The $\lambda/2$ plate is used to rotate the photons linear polarization and
 is placed in front of a spatial light modulator (SLM) because the modulation of the SLM depends on polarization of the incident photon \cite{wander,Moreno}. Considering
the pump beam direction as the $z$-direction, the triple-slit plane is in the $x-y$ plane with the smaller dimension in the $x$-direction. The slits's width are equal
to 2a=100 $\mu$m and are separated from each other by $2d'=250$ $\mu$m. A 300 mm focal length lens $L$, placed 50 mm before the crystal is used for focusing the pump beam at the
triple-slit plane such that an entangled two-photon state in transverse path variables is generated after the triple-slit \cite{Mikami}. Let's assume the pump beam profile is narrow in the x-direction at the slits's plane such that it can be approximated by a delta function. The predicted photon pair state in this case is
\begin{equation}
 |\psi_{th}\rangle = \frac{1}{\sqrt{3}}(e^{i\phi}|0\rangle_{s}|2\rangle_{i} + |1\rangle_{s}|1\rangle_{i} + e^{i\phi}|2\rangle_{s}|0\rangle_{i}), \label{state}
\end{equation}
where $\{|0 \rangle,|1\rangle,|2 \rangle\}$ are the photon path states provided by the slits, $\phi$ is a phase that depends on the path length difference between the slits and the crystal center, in our case $\phi$=1.94 rad, and $s$ and $i$ correspond to signal and idler, respectively \cite{Neves3}. As it is shown theoretically in reference \cite{Neves3} and measured in \cite{trans1}, the pump profile at the triple-slit plane determines the photons spatial correlation at the slits and therefore, the entanglement of the transmitted photon pair quantum state in path variables. Pump beam transverse profile
 narrower than the slits separation at the triple slit plane is required for generating a maximum entangled state. In this condition signal and idler always cross opposite slits.

The SLM is positioned just behind the triple-slit, at 2.0 mm from it, for avoiding diffraction. Signal and idler photons are reflected by the SLM, returning through the slits's paths.
The SLM is a Holoeye Photonics LC-R 2500 (reflecting type SLM), which has 1024 x 768 pixel resolution (each pixel consists in a 19 x 19 $\mu$m square) and it is controlled
by a computer.   Polarizers, P$_{s}$ and P$_{i}$, are
placed just before the detectors because the SLM adds path phase to the reflected photons and modifies their state polarization. The required photon modulation (phase and amplitude) is obtained by choosing specific SLM grey scales and output polarizers rotation angles \cite{wander,Moreno}.

Two 200 mm focal length lenses, L$_{s1}$ and L$_{i1}$, are placed at the focal distance from the detectors, to generate an interference pattern on
the detectors' planes. On the other hand, triple-slit image at the detectors'plane can be obtained by replacing L$_{s1}$ and L$_{i1}$ lenses by  125 mm focal length lenses, L$_{s2}$ and L$_{i2}$, placed at distance 2f=250 mm after the triple-slit in the 2f-2f configuration.
All measurements necessary to reconstruct the density operator $\rho$
are made in the Fourier plane. By detecting at $x_{i}=0$ or $x_{s}=0$ we are able to implement a positive operator that is
proportional to the projector $P=| + \rangle \langle + | $, with $| + \rangle = \frac{1}{\sqrt{3}}( e^{i\phi/2}|0 \rangle + |1\rangle + e^{i\phi/2}|2 \rangle)$ where $\phi/2=0.81$ rad.
100 $\mu$m single slits, S$_{s}$ and S$_{i}$, are placed in front of the detectors. The smallest dimension, of each slit, is
parallel to the corresponding x-direction. Signal and idler beams were focused on the detectors with a microscope objective lens and two
interference filters, centered at 810 nm, with 10 nm Full Width at Half Maximum, were kept before the objective lenses. Pulses from the detectors are
sent to a photon-counter and a coincidence detection setup with 5.0 ns resolving time.

Fig. \ref{montagem}-(b) shows the state preparation part of the experimental setup. The transmitted photon pairs through the triple slit are in a state
close to the state shown in Eq. \ref{state}. This was demonstrated theoretically in reference \cite{Neves3}. A lens before the crystal focus the laser beam at the triple-slit plane such that its transversal profile is narrower than the slits separation. A dicroic mirror placed after the crystal, not shown in the Figure, reflects the pump beam and transmits the photon pairs. Fig. \ref{montagem}-(c) shows the state determination part of the setup.
The SLM reflects the photons pairs through the same slits'paths toward the signal and idler detectors. Photons at the Fourier Plane, i. e., at the focal plane of lenses $L_i$ and $L_s$. Two polarizers $P_s$ and $P_i$ are placed at the photon paths. Different polarizer rotation angles combined with specific grey scales introduce the necessary phase and attenuation to the photon paths \cite{Moreno,wander}.

\section{Experimental strategy to implement the measurement operators using a SLM}

\begin{figure}[!htbp]
		\includegraphics[width=9cm]{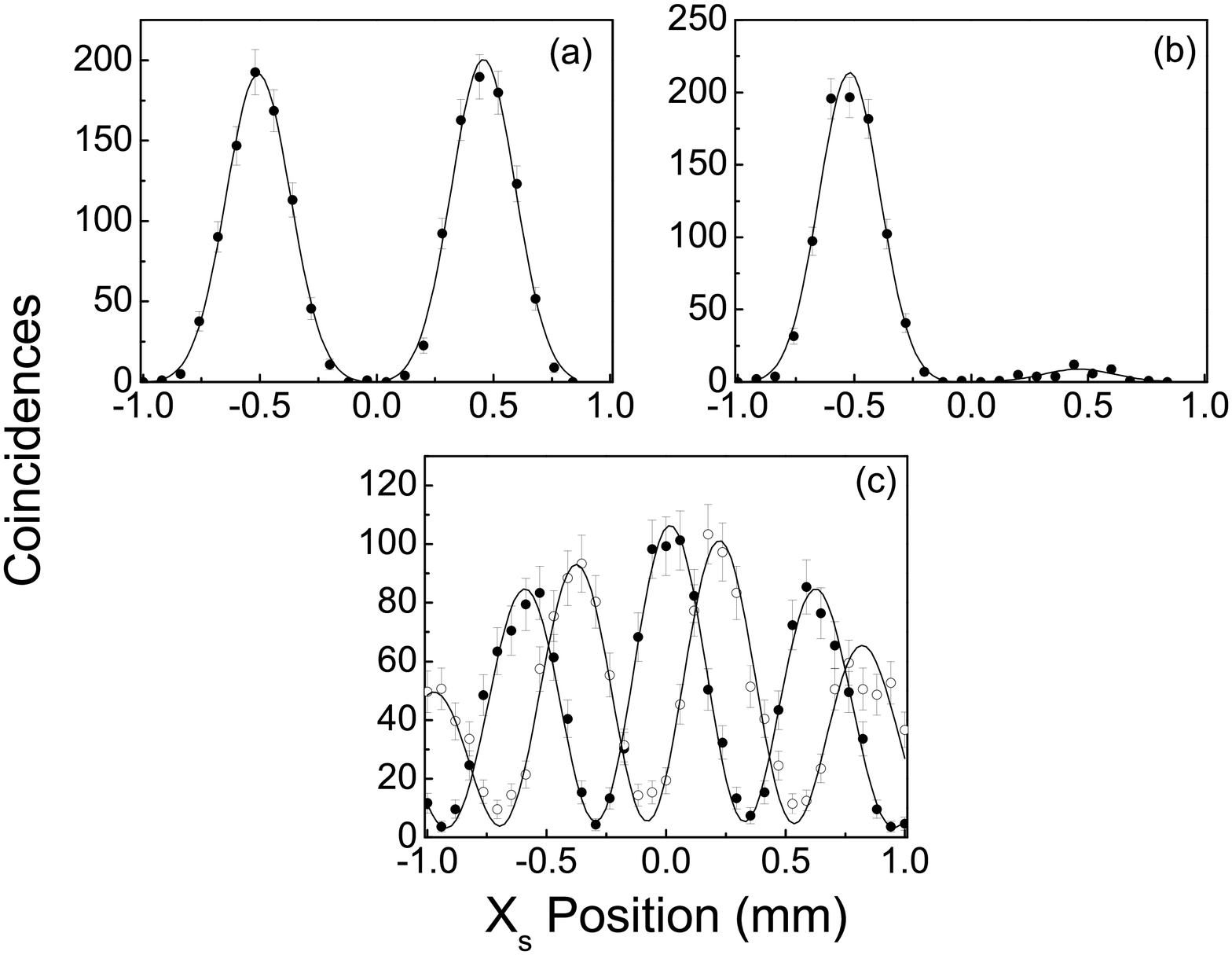}
		\vspace{0.0cm}
		\caption{\small{Double-slit images and interference patterns are shown for demonstrating the ability of the SLM to block a photon paths and to add relative phases between their paths. In (a) is showed the double-slit image with two different gray scales in each slit which gives a phase difference of $2\pi/3$ without attenuation. In (b) is showed the double-slit image with the gray scales that attenuate as much
		as possible one slit image in relation to the other. In (c) are showed double-slit interference patterns. The solid circles represent the pattern with the
		SLM turned off while the open circles represent the pattern with the same gray scales from the image (a). The image measurements were made in 10
		seconds while the pattern measurements were made in 60 seconds.}}
	\label{ponto}
\end{figure}

In the Equations \ref{1} and \ref{2} were defined the POVM $\{\hat{\Pi}_{ij}\}$($i,j=1,..,9$) which we use to reconstruct the two-qutrit density operator with the minimum number of measurements. To
prepare the measurement operators we must be able to modify the state amplitude and phase of the photons transmitted by each slit. It was shown in \cite{Moreno} that the phase and amplitude
modulation of this SLM depend on input and output polarizers angle orientation. The liquid crystal display was divided into
six regions, each with its own gray scale. Each region corresponding to a slit. Each region attenuates and/or adds phase to the photon path state defined by the slits.
By adjusting the $\lambda/2$ wave plate angle, and the angle orientation of the output polarizers, $P_{i}$ and $P_{s}$,
we are able to obtain the relative path modulation. To implement the measurement operators it was necessary to generate a phase difference of $2\pi/3$ between any
two slits and completely attenuate the third one (Eq. \ref{1} and Eq. \ref{2}). In Fig. \ref{ponto}, we show experimentally that we are able to perform that kind
of control. We substitute for a moment the triple-slit by a double-slit because it is easier to measure the spatial relative phase added by the SLM with a
double-slit than with a triple-slit interference pattern.
Fig. \ref{ponto} shows the images and two-slit interference patterns with gray scales at the SLM used for the
tomography. The images showed in Figs. \ref{ponto}(a) and \ref{ponto}(b) were done by keeping the detector D$_{i}$ fixed at $x_{i}=0$ (and using $L_{i1}$ lens) while the detector D$_{s}$ was scanned in the x direction (and using the $L_{s2}$ lens). The SLM screen part that reflects the idler photon was kept with
the same gray scale in both slits. The other part of the SLM, that reflects the signal photon was divided into two regions with diferent gray
scales. Fig. \ref{ponto}(a) shows the image of the slits with the gray scales that give phase difference of $2\pi/3$ without any relative attenuation. Using a gaussian
fit, it was obtained an area of A$_{1}$=(68$\pm$2) and A$_{2}$=(72$\pm$2) in arbitrary units  for the left and right peaks, showing that these two gray scales do not attenuate relatively the reflected photon flux. Fig. \ref{ponto}(b) shows the double-slit image when is used gray scales that attenuate the reflected photon flux through only one of the slits. A$_{1}$=(70$\pm$2) and A$_{2}$=(3.3$\pm$0.5) in arbitrary units are the areas of the left and right peaks showing 96$\%$ of attenuation. For measuring the interference pattern, the detector D$_{i}$ was kept at $x_{i}=0$ and it was used the lens $L_{i2}$. The part of the SLM screen that reflects the idler photon was kept with the same gray scale in both slits. Using the $L_{s1}$ lens and scanning the Detector D$_{s}$ in x direction we measured
the interference pattern with (opened circles) and without (closed circles) the SLM modulation (Fig. \ref{ponto}(c)). By a theoretical fit, it was obtained the phase difference between the two interference patterns: $\Delta\phi$=(2.10$\pm$0.04) rad ($\approx 2\pi/3$ rad). The previous characterization of the SLM shows that we can generate in a controlled way the required phase difference and slit attenuation with great accuracy. Therefore, we have all the necessary tools to
implement the minimum two-qutrit tomography using the one-qutrit SIC-POVM.

\section{Experimental generation of two-qutrit quantum state and the minimum quantum tomography}

In order to prepare  the transversal path entangled two-qutrit state, the pump beam was focused at the triple-slits plane, using the lens $L$ with 300 mm focal distance,
as showed in Fig.\ref{montagem}. Triple-slit images were measured for testing the state preparation. Like the previous image measurement, the
detector D$_{i}$ was kept fixed at the image of one of the slits while the detector D$_{s}$ was scanned in $x$ direction. At the Fourier plane, interference patterns were measured by scanning the detector $D_{s}$ while keeping $D_{i}$ fixed at $x_{i}=0$mm and $x_{i}=350$mm.  This image correlation together with the measured conditional fringes in the interference patterns are entanglement signatures
\cite{Rossi,Greenberger,Fonseca,Neves4}. The triple-slit images and conditional interference patterns were obtained when the SLM was turned off.

Eqs. (\ref{1}) and (\ref{2}) show the states that construct the measurement operators $\Pi_{i}=|\psi_{i} \rangle \langle \psi_{i} |$ with $|\psi_{i} \rangle= e^{i\alpha_{i}}| j \rangle + e^{-i\alpha'_{i}} |k \rangle$, $\alpha_{i},\alpha'_{i}=\{0,2\pi/3,-2\pi/3\}$ and $j,k=0,1,2$ ($j<k$).
Experimental implementation of these measurement operators requires to detect a photon that traveled through two different paths (coming from slit j or k) with a relative path phase difference of 2$\alpha$. This phase difference is added by the SLM. The third path from a third slit not present in the measurement operator
$\Pi_{i}$ is eliminated by rotating the reflected photon polarization with a suitable gray scale at the SLM and by "blocking" the photon with polarizers $P_{s}$ or $P_{i}$.

\begin{figure}[!htbp]
		\includegraphics[width=8.5cm]{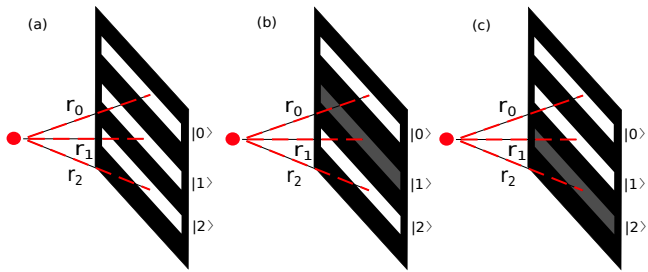}
		\vspace{0.0cm}
		\caption{\small{(Color online) Extra phase gained by photons signal or idler to the different path length from each slit to the detectors ($r_{0}=r_{2}\neq r_{1}$). In (a), we see the possible paths which a photon can take
		from each slit. Attenuating the middle slit, as it is shown in (b), means that the path phase gained by the photon is the same by passing through the upper or the lower slit, and this does not affect any possible operator implementation.  In (c) the lower slit is blocked and photons that cross the remained slits will have different path phases. In this case, this extra phase has to be considered in the measurement operator implementation.}}\label{fases}
\end{figure}

The path phase operation can be written as an unitary matrix
$U=U_{i}\otimes U_{s}$ where
\begin{equation}
U_{e} = \left(
\begin{array}{ccc}
e^{i\phi_{e}} & 0 & 0 \\
0 & 1 & 0\\
0 & 0 & e^{i\phi_{e}} \\
\end{array}
\right),\\
\end{equation}
with $e=i,s$ and $\phi_{e}$ is a phase due the photon's path until each slit. Adding this matrix to the POVM we see that $U^{\dagger}\Pi_{ij}U=\Pi'_{ij}$, $\Pi'_{i,j}$ still continues being a POVM unitarily equivalent to the previous one.
This $\phi_{e}$ phase was obtained by a theoretical fit of the conditional interference patterns with one of the photon paths attenuated. The obtained value is
$\phi_{e}=0.922$ rad.

\begin{figure}
	\includegraphics[width=6cm]{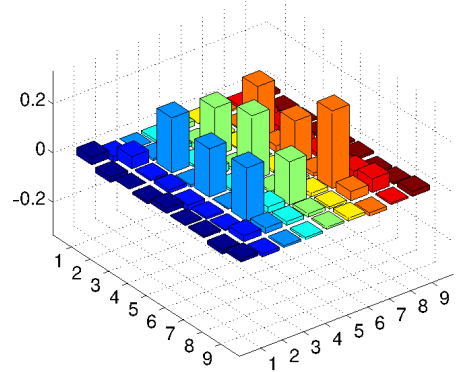}
\caption{(Color online) Real part of the density operator measured for a two-qutrit state.} \label{rore}
\end{figure} 

\begin{figure}
	\includegraphics[width=6cm]{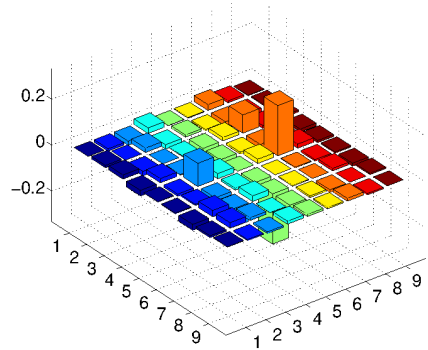}
\caption{(Color online) Imaginary part of the density operator measured for a two-qutrit state.} \label{roima}
\end{figure}

\begin{widetext}

\begin{figure}[H]

\begin{equation}
\small
\begin{split}
 \hat{\rho}_{tom}^{Re}=\left(
 \begin{array}{ccccccccc}
0.04   &  -0.02 & 0.01  & 0.00  & 0.03  & -0.01& 0.02  & -0.03 & 0.01\\
-0.02  &  0.02  & 0.00 & 0.01  & -0.01 & 0.00 & -0.01 & 0.00  & -0.02\\
0.01   &  0.00 & 0.18  & 0.01  & 0.18  & 0.03 & 0.21  & 0.04  & 0.00\\
0.00   &  0.01  & 0.01  & 0.04  & 0.01  & 0.00 & 0.05  & 0.01  & -0.01\\
0.03   &  -0.01 & 0.18  & 0.01  & 0.24  & 0.03 & 0.18  & 0.03  & 0.01\\
-0.01  &  0.00  & 0.03  & 0.00  & 0.03  & 0.02 & 0.04  & 0.03  & 0.01\\
0.02   &  -0.01 & 0.21  & 0.05  & 0.18  & 0.04 & 0.34  & 0.05  & 0.02\\
-0.03  &  0.00  & 0.04  & 0.01  & 0.03  & 0.03 & 0.05  & 0.10  & 0.03\\
0.01   &  -0.02 & 0.00  & -0.01 & 0.01  & 0.01 & 0.02  & 0.03  & 0.03\\
\end{array}
\right),  \label{rhoreal} \\
\end{split}
\end{equation}
\\

\begin{equation}
\small
\begin{split}
 \hat{\rho}_{tom}^{Im}=\left(
 \begin{array}{ccccccccc}
0.00   &  0.00 & 0.00  & -0.03  & -0.01  & 0.00 & -0.04  & -0.02 & -0.01\\
0.00  &  0.00 & -0.01  &  0.02  &  0.00 &  0.01 &  0.02  & 0.02  & 0.01\\
0.00   &  0.01 & 0.00   & 0.00   &  0.09  &  0.00 & -0.09  & 0.00 & -0.01\\
0.03   & -0.02 &  0.00  & 0.00   &  0.03  &  0.01 &  0.02  & -0.02 & 0.02\\
0.01   &  0.00 & -0.09  & -0.03  &  0.00  & -0.02 & -0.22  & -0.03 & -0.01\\
0.00   & -0.01 & 0.00  & -0.01  &  0.02  & 0.00  & -0.02  &  0.02 & 0.01\\
0.04   & -0.02 &  0.09  & -0.02  &  0.22  &  0.02 &  0.00  & 0.00  & 0.02\\
0.02   & -0.02 &  0.00  &  0.02  &  0.03  & -0.02 &  0.00  & 0.00  & 0.00\\
0.01   & -0.01 &  0.01  &  -0.02 &  0.01  & -0.01 & -0.02  & 0.00  & 0.00\\
\end{array}
\right).  \label{rhoimag} \\
\end{split}
\end{equation}
\end{figure}
\end{widetext}

With the correct gray scales at the SLM, that give us the desired attenuations/phases, we are able to implement the 81 measurement operators
$\{\Pi_{i,j}\}$ and we obtain the tomographic reconstruction of the state. Signal and idler detectors were kept at $x=0$ in all measurements needed for the tomography reconstruction of the density operator. Fig. \ref{fases}(a) shows the different paths from the slits to the detector signal or idler ($r_{0}=r_{2}\neq r_{1}$). In Fig. \ref{fases}(b), the middle slit is blocked and the relative phase between the different photon paths is zero. In Fig. \ref{fases}(c), the lower slit is blocked. In this case a relative phase appears between the remaining paths $\phi_{e}=\frac{2\pi}{\lambda}(r_{0}-r_{1})$ ($e=i,s$). The density operator is written in the two-qutrit logical bases, i.e., $\{|00\rangle,|01\rangle,|02\rangle,|10\rangle,|11\rangle,
|12\rangle,|20\rangle,|21\rangle,|22\rangle\}$ basis. Measurements were taken in an acquisition time of 1500 seconds. The real and imaginary parts of the obtained density operator ($\rho_{tom}=\rho_{tom}^{Re}+i\rho_{tom}^{Im}$) are showed in equations (\ref{rhoreal}) and (\ref{rhoimag}). Moreover, Figures (\ref{rore}) and (\ref{roima}) show the histograms of the real and imaginary parts of the density operator obtained. The peaks that appear in the histogram of imaginary part may be due to the Gouy phase,  a differing phase that modes of different order
pick up on propagation. This longitudinal phase appears because of a non-optimal position between the Fourier lenses $(L_{s1},L_{i1})$ and the detectors. Similar behavior was observed in \cite{agnew}.

\section{Discussion}

As mentioned above we measured the conditional images of the triple slit by coincidence detection between signal and idler slit images at the image plane \cite{trans1,wander}. Detector $D_{s}$ is scanned while detector $D_{i}$ is kept fixed at one of the three idler slit images. Three plots of the Coincidence counts in terms of the $D_{s}$ detector positions were made for the three fixed $D_{i}$ positions. The areas below the peaks in the graphs, corresponding to the photon path correlation after crossing the different slits, were calculated. These areas were normalized by dividing each one by the sum of the areas.

\begin{table}
\caption{Normalized areas under the image curves and diagonal elements from the reconstructed density operator.} \label{tabela}
\begin{center}
\begin{tabular}{c|c|c}
\hline
\hline
\hspace{2mm}\textbf{States}\hspace{2mm}  & \hspace{2mm}\textbf{Norm. area under}\hspace{2mm}& \hspace{2mm}\textbf{Diagonal elements}\hspace{2mm} \\
        &  \textbf{curve} \\
\hline
\hline
$|0\rangle_{s}|0\rangle_{i}$ & 0   & 0.04\\
$|0\rangle_{s}|1\rangle_{i}$ & 0.02  & 0.02\\
$|0\rangle_{s}|2\rangle_{i}$ & 0.23 & 0.18\\
$|1\rangle_{s}|0\rangle_{i}$ & 0.01   & 0.04\\
$|1\rangle_{s}|1\rangle_{i}$ & 0.36 & 0.24\\
$|1\rangle_{s}|2\rangle_{i}$ & 0.01  & 0.02\\
$|2\rangle_{s}|0\rangle_{i}$ & 0.32 & 0.34\\
$|2\rangle_{s}|1\rangle_{i}$ & 0.02  &  0.10\\
$|2\rangle_{s}|2\rangle_{i}$ & 0   & 0.03\\
\hline
\hline
\end{tabular}
\end{center}
\end{table}

Table \ref{tabela} shows the normalized areas under each peak which correspond to the population of each two-qutrit state and also the respective population obtained from the minimum tomography method (diagonal elements). We see that the generated state is correlated, in a way that if the
idler photon passes through one of the slits, the signal photon will pass through the symmetrically opposite one \cite{trans1,Neves3}. 

We have checked the quality of the reconstructed density operator by comparing the theoretical state that we desired to generate (Eq. \ref{state}) with the expected values obtained from the measured density operator ($\rho_{tom}$). The fidelity of the measured $\rho$ with respect to the theoretically predicted state with a varying phase was calculated. We obtained a maximum value of 0.86 for a phase of $\phi$=0.58 rad different from the initially predicted $\phi$=1.94 rad. As mentioned above, this phase difference may be due a longitudinal phase that appears because of a non-optimal position between the Fourier lenses ($L_{s1}$,$L_{s1}$) and the detectors \cite{james}.

\begin{figure}[!htbp]
		\includegraphics[width=9cm]{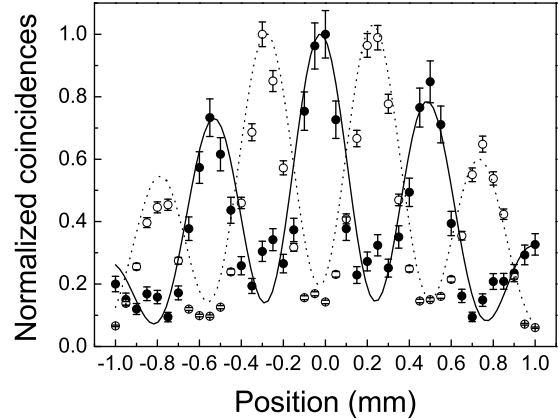}
		\vspace{0.0cm}
		\caption{\small{Comparison between the interference pattern obtained from tomography (continuous and dashed curves) with the ones measured at the Fourier plane (closed and opened circles). In the closed circles, Idler detector is kept fixed at x$_{i}$ = 0 and the opened open circles the idler detector is kept fixed at $x_{i}=350\mu m$. Error bars in the Figure are calculated from the statistical fluctuation (Poisson Distribution) of the coincidence counts.}}\label{fit}
\end{figure}

We can also test the quality of the reconstructed density operator by comparing the predicted interference pattern $IP(x)$ using the tomographed $\rho_{tom}$ for this calculation with the independent measured interference pattern at the Fourier plane.
The predicted interference pattern due to the density operator is obtained from
\begin{equation}
 IP(x_{i},x_{s})=Tr[\Pi_{i}(x_{i})\Pi_{s}(x_{s})\rho_{tom}], \label{IP}
\end{equation}
where $\Pi_{j}(x)$ (j=1,2) represent the measurement operator from the triple slits to the Fourier plane. The operator $\Pi_{j}(x)$ in Eq. \ref{IP} is given by
\begin{equation}
 \Pi_{j}(x_{j})=\int_{x_{j}-b}^{x_{j}+b}sinc^{2}\big(\frac{kax'}{2f}\big)\sum_{l,m}e^{\frac{ikd(l-m)x'}{2f}} |l\rangle \langle m|dx' \label{projector}
\end{equation}
where $2b$, $k$ and $f$ are the slit aperture width in front of the detector j ($j=i,s$), the pump beam wave number and the focal length of $L_{i1}$($L_{s1}$) lens. The sum is over all slits $l, m (l, m= 0, 1, 2)$. Using the above equations we could obtain the interference pattern and finally compare with a measured interference pattern. Figure \ref{fit} shows that the interference patterns obtained from tomography is closer to the ones measured at the Fourier plane for the same prepared state. Error bars in the Fig. \ref{fit} are calculated from the statistical fluctuation (Poisson Distribution) of the coincidence counts.

\begin{figure}[!htbp]
		\includegraphics[width=8.5cm]{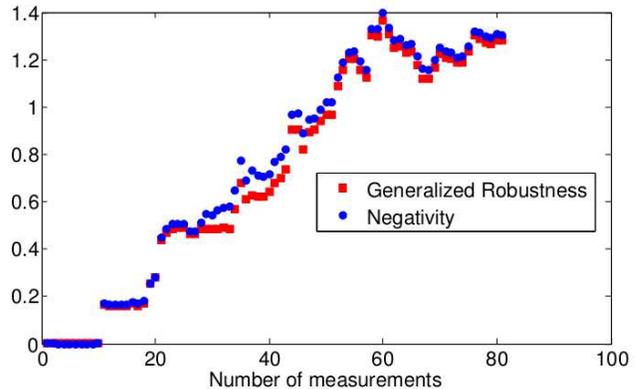}
		\vspace{0.0cm}
		\caption{\small{(Color online) Dependence between the entanglement quantifiers and the number of measurements used for the reconstruction of two-qutrit density operator. The quantifiers
		used were the robustness and negativity. For a two-qutrit state these two quantifiers have the maximum value of 2.}}\label{tomografia_3x3}
\end{figure}

We quantified the entanglement of the two-qutrit state by calculating the robustness and negativity \cite{wei} (Fig. \ref{tomografia_3x3}) . These entanglement quantities were calculated from the reconstructed density operator with partial information. The Variational Tomography Method \cite{Maciel1,Maciel2} allows us to obtain a density operator with a certain number of POVM elements smaller than $d^{4}=81$. For each obtained density operator, robustness and negativity were calculated. Fig. \ref{tomografia_3x3} shows the two entanglement measurements calculated from $\rho_{tom}$ reconstructed from $N<d^{4}$ POVM elements. Note that entanglement is already detected by using $N>11$ POVM elements in the reconstruction and for $N>60$ entanglement measurement is stabilized around 1.3 (maximum value=2 for two-qutrits).

\section{Conclusion}

In this work, we experimentally performed the two-qutrit minimum quantum state tomography using one-qutrit symmetric informationally complete positive operator-valued measure. This work was done by using two photonic qutrit states prepared in the transverse path degree of freedom
when photon pairs generated by spontaneous parametric down conversion cross a triple-slit.Measurements for
the tomography were made only performing changes in gray scales of the SLM  to provide phase shift and/or attenuation keeping the detectors fixed at the same position. The quality of tomography was tested by comparing the predicted slit conditional images and conditional interference patterns with independent measurements.
Entanglement of the two-qutrit system was measured via the negativity and generalized robustness, calculated from the reconstructed density operator obtained from $N<d^{4}$ POVM elements. This complete characterization of the two-qutrit system via a minimum quantum state tomography scheme and the variational tomography method can be generalized to higher dimension system where the POVM is calculated.

\section{Acknowledgements}

This work is part of the Brazilian National Institute for Science and Technology for Quantum Information and was supported by the Brazilian agencies CNPq, CAPES, and FAPEMIG. We acknowledge the \textit{EnLight} group for very useful discussions. A. D. acknowledges support from MSI P10-030F.


\end{document}